%% file: arxiv.tex
\documentclass[graybox]{svmult}

\usepackage{mathptmx}       
\usepackage{helvet}         
\usepackage{courier}        
\usepackage{type1cm}        
\usepackage{makeidx}         
\usepackage{graphicx}        
\usepackage{multicol}        
\usepackage[bottom]{footmisc}

\makeindex             

\include{ad2012Packages}\usepackage[round]{natbib}\renewcommand{\cite}[1]{\citep{#1}} 
\include{ad2012NewCommands} 

\include{newCommands} 

\begin{document}

{\include{paper}}

\end{document}

%% file: ad2012Packages.tex
\usepackage{amsmath,amsfonts}
\usepackage{listings}
\usepackage{url}
\usepackage[dvips]{epsfig,psfrag}
\usepackage{booktabs}
\usepackage{paralist}

%% file: ad2012NewCommands.tex
\newcommand{\reffig}[1]{{Fig.~\ref{#1}}}
\newcommand{\brefsec}[1]{{Section~\ref{#1}}}

\newcommand{\refeqn}[1]{{(\ref{#1})}}

%% file: newCommands.tex
\newcommand{\PaperXXlangLong}{Forward And Reverse Fortran Extension Language}
\newcommand{\PaperXXlang}{\textsc{Farfel}}
\newcommand{\PaperXXpreprocessor}{\textsc{Farfallen}}
\newcommand{\PaperXXfort}{\textsc{Fortran}}
\newcommand{\PaperXXalgol}{\textsc{Algol}}

\newcommand{\PaperXXadifor}{\textsc{Adifor}}
\newcommand{\PaperXXtapenade}{\textsc{Tapenade}}

\newcommand{\PaperXXadic}{\textsc{adic}}
\newcommand{\PaperXXClang}{\textsc{c}}
\newcommand{\PaperXXGNU}{\textsc{gnu}}
\newcommand{\PaperXXGCC}{\textsc{gcc}}
\newcommand{\PaperXXVLAD}{\textsc{vlad}}
\newcommand{\PaperXXStalingrad}{\textsc{Stalingrad}}

\newcommand{\PaperXXpar}[1]{%
\mbox{}
\par
\noindent
\textbf{#1} \\
}

\DeclareMathOperator*{\argmax}{argmax}

\definecolor{darkblue}{rgb}{0,0,0.7}
\definecolor{darkgreen}{rgb}{0,0.5,0}
\definecolor{darkred}{rgb}{0.7,0,0}

%% file: paper.tex
\title*{AD \emph{in} Fortran}
\subtitle{Part 1: Design}
\titlerunning{AD \emph{in} Fortran: Design}
\author{Alexey Radul \and Barak A. Pearlmutter \and Jeffrey Mark Siskind}
\institute{
Alexey Radul \at Hamilton Institute, National University of Ireland
Maynooth, Ireland, \url{alexy.radul@nuim.ie}
\and
Barak A. Pearlmutter \at Department and Computer Science and Hamilton Institute, National University of Ireland
Maynooth, Ireland, \url{barak@cs.nuim.ie}
\and
Jeffrey Mark Siskind \at Electrical and Computer Engineering, Purdue
University, IN, USA, \url{qobi@purdue.edu}}
\maketitle

\abstract{We propose extensions to \PaperXXfort\ which integrate forward and
  reverse Automatic Differentiation (AD) directly into the programming model.
  Irrespective of
  implementation technology, embedding
  AD constructs directly into the language
  extends the reach and convenience of AD while
  allowing abstraction of concepts of interest to scientific-computing
  practice, such as root finding, optimization, and finding equilibria
  of continuous games.  Multiple different subprograms for these tasks
  can share common interfaces, regardless of whether and how they use
  AD internally.  A programmer can maximize a function \lstinline{F}
  by calling a library maximizer, \lstinline{XSTAR=ARGMAX(F, X0)},
  which internally constructs derivatives of \lstinline{F} by AD,
  without having to learn how to use any particular AD tool.  We
  illustrate the utility of these extensions by example: programs
  become much more concise and closer to traditional mathematical
  notation.  A companion paper describes how these extensions
  can be implemented by a
  program that generates input to existing \PaperXXfort-based AD tools.}
\keywords{Nesting, multiple transformation, forward mode, reverse
  mode, \PaperXXtapenade, \PaperXXadifor, programming-language
  design}

\section{Introduction}\label{PaperXXsec:intro}

The invention of \PaperXXfort\ was a major advance for numeric computing,
allowing
\begin{equation*}
  g(x;\alpha,\beta) = \frac{\beta^{\alpha}}{\Gamma(\alpha)} \,
  x^{\alpha-1} \, e^{-\beta x}
\end{equation*}
to be transcribed into a natural but unambiguous notation
\begin{lstlisting}
      FUNCTION G(X,ALPHA,BETA)
      G=BETA**ALPHA/GAMMA(ALPHA)*X**(ALPHA-1)*EXP(-BETA*X)
      END
\end{lstlisting}
which could be automatically translated into an executable program.
However, transcribing
\begin{equation*}
  x_{i+1} = x_{i} - f(x_{i})/f'(x_{i})
\end{equation*}
to \PaperXXfort\ in
\begin{lstlisting}
      FUNCTION RAPHSN(F, FPRIME, X0, N)
      EXTERNAL F, FPRIME
      X = X0
      DO 1690 I=1,N
 1690 X = X-F(X)/FPRIME(X)
      RAPHSN = X
      END
\end{lstlisting}
requires that the \emph{caller} provide both \lstinline{F} and
\lstinline{FPRIME}.
Manually coding the latter from the former is, in most cases, a
mechanical process, but tedious and error prone.

This problem has traditionally been addressed by arranging for an AD
preprocessor to produce \lstinline{FPRIME} \cite{Speelpenning1980CFP,
  Wengert1964ASA}.
That breakthrough technology not only relieves the programmer of the
burden of mechanical coding of derivative-calculation codes, it also
allows the derivative code to be updated automatically, ensuring
consistency and correctness.
However, this \emph{caller derives} discipline
has several practical difficulties.
First, the user must learn how to use
the AD preprocessor, which constitutes a surprisingly serious barrier
to adoption.
Second, it makes it very difficult to experiment with the use of
different sorts of derivatives (e.g., adding a Hessian-vector product step
in an optimization) in such called subprograms, or to experiment with
different AD preprocessors.
Third, although preprocessors might be able to process code which has
already been processed in order to implement nested derivatives, the
maneuvers required by current tools can be somewhat arcane
\cite{Siskind2008PtA}.
Fourth, software engineering principles of locality
and atomicity are being violated:
knowledge of what derivatives are needed is
distributed in a number of locations which must be kept consistent;
and redundant information, which must also be kept consistent, is
being passed, often down a long call chain.
We attempt to solve these problems, making the use of AD
more concise, convenient, and intuitive to the scientific
programmer, while keeping to the spirit of \PaperXXfort.
This is done using the \emph{\PaperXXlangLong} or \PaperXXlang,
a small set of extensions to \PaperXXfort,
in concert with an implementation strategy which leverages existing
\PaperXXfort\ compilers and AD preprocessors
\cite{Bischof1992AGD, Hascoet2004TUG}.

The remainder of the paper is organized as follows:
\brefsec{PaperXXsec:language} describes \PaperXXlang.
\brefsec{PaperXXsec:example} describes a concrete example \PaperXXlang\
program to both motivate and illuminate the proposed extensions.
\brefsec{PaperXXsec:discuss} situates this work in its broader context.
\brefsec{PaperXXsec:conclusion} summarizes this work's contributions.
A companion paper \cite{PaperXXimplementation} describes how \PaperXXlang\ can
be implemented by generating input to existing AD tools.

\section{Language Extensions}\label{PaperXXsec:language}

\PaperXXlang\ consists of two principal extensions to \PaperXXfort: syntax for
AD and for nested subprograms.
We currently support only \PaperXXfort77, but there is no barrier,
in principle, to adding \PaperXXlang\ to more recent dialects.

\PaperXXpar{Extension 1: AD Syntax}
Traditional mathematical notation allows one to specify
\begin{equation*}
 \phi' = \frac{\textrm{d}}{\textrm{d}\sigma} \Bigl(
  \frac{1}{\sqrt{2\pi\sigma^2}}
  \exp -\frac{1}{2}\Bigl(\frac{x-\bar{x}}{\sigma}\Bigr)^2
\Bigr)
\end{equation*}
By analogy, we extend \PaperXXfort\ to encode this as
\begin{lstlisting}
      ADF(TANGENT(SIGMA) = 1)
      PHI = 1/SQRT(2*PI*SIGMA**2)*EXP(-0.5*((X-XBAR)/SIGMA)**2)
      END ADF(PHIPRM = TANGENT(PHI))
\end{lstlisting}
which computes the derivative \lstinline{PHIPRM} of \lstinline{PHI}
with respect to \lstinline{SIGMA} by forward AD.\@
For syntactic details see companion paper~\cite{PaperXXimplementation}.

An analogous \PaperXXlang\  construct supports computing
the same derivative with reverse AD:
\begin{lstlisting}
      ADR(COTANGENT(PHI) = 1)
      PHI = 1/SQRT(2*PI*SIGMA**2)*EXP(-0.5*((X-XBAR)/SIGMA)**2)
      END ADR(PHIPRM = COTANGENT(SIGMA))
\end{lstlisting}
Note that with the \lstinline{ADR} construct, the \emph{dependent} variable
appears at the beginning of the block and the \emph{independent} variable at
the end---the variables and assignments in the opening and closing statements
specify the desired inputs to and outputs from the reverse phase, whereas the
statements inside the block give the forward phase.

These
constructs allow not just convenient expression of AD, but also
modularity and encapsulation of code which employs AD.\@
For instance, we can write a general scalar-derivative subprogram
\lstinline{DERIV1} at user level
\begin{lstlisting}
      FUNCTION DERIV1(F, X)
      EXTERNAL F
      ADF(X)
      Y = F(X)
      END ADF(DERIV1 = TANGENT(Y))
      END
\end{lstlisting}
which could be used in, for example,
\begin{lstlisting}
      FUNCTION PHI(SIGMA)
      PHI = 1/SQRT(2*PI*SIGMA**2)*EXP(-0.5*((X-XBAR)/SIGMA)**2)
      END
      PHIPRM = DERIV1(PHI, SIGMA)
\end{lstlisting}
\lstinline{DERIV1} can be changed to
use reverse AD without changing its API:
\begin{lstlisting}
      FUNCTION DERIV1(F, X)
      EXTERNAL F
      ADR(Y)
      Y = F(X)
      END ADR(DERIV1 = COTANGENT(X))
      END
\end{lstlisting}
allowing codes written with \lstinline{DERIV1} to readily switch between using
forward and reverse AD.

To take a more elaborate example, we can write a general gradient
calculation \lstinline{GRAD} using repeated forward AD:
\begin{lstlisting}
      SUBROUTINE GRAD(F, X, N, DX)
      EXTERNAL F
      DO 1492 I=1,N
      ADF(TANGENT(X(J)) = 1-MIN0(IABS(I-J),1), J=1,N)
      Y = F(X)
 1492 END ADF(DX(I) = TANGENT(Y))
      END
\end{lstlisting}
(Note that the \lstinline{ADF} and \lstinline{ADR} constructs support
implied-\lstinline{DO} syntax for arrays.)

This can be modified to instead use reverse AD without changing the API:
\begin{lstlisting}
      SUBROUTINE GRAD(F, X, N, DX)
      EXTERNAL F
      ADR(Y)
      Y = F(X)
      END ADR(DX(I) = COTANGENT(X(I)), I=1,N)
      END
\end{lstlisting}
Although not intended to support checkpoint-reverse AD, our constructs
are sufficiently powerful to express a reverse checkpoint:
\begin{lstlisting}
C      CHECKPOINT REVERSE F->G.  BOTH 1ST ARG IN, 2ND ARG OUT
       CALL F(X, Y)
       ADR(COTANGENT(Z(I)) = $\ldots$, I=1,NZ)
       CALL G(Y, Z)
       END ADR(DY(I) = COTANGENT(Y(I)), I=1,NY)
       ADR(COTANGENT(Y(I)) = DY(I), I=1,NY)
       CALL F(X, Y)
       END ADR(DX(I) = COTANGENT(X(I)), I=1,NX)
\end{lstlisting}

This sort of encapsulation empowers numeric programmers to
conveniently experiment with the choice of differentiation method, or
with the use of various sorts of derivatives, including higher-order
derivatives, without tedious modification of lengthy call chains.

\PaperXXpar{Extension 2: Nested Subprograms}
We borrow from \PaperXXalgol~60 \cite{PaperXXNaur-1963xxx} and
generalize the \PaperXXfort\ ``statement function'' construct by
allowing subprograms to be defined inside other subprograms, with
lexical scope.

For example, given a univariate maximizer \lstinline{ARGMAX}, we can express
the idea of line search as follows:
\pagebreak[4]
\begin{lstlisting}
C     MAXIMIZE F ALONG THE LINE PARALLEL TO XDIR THROUGH X
      SUBROUTINE LINMAX(F, X, XDIR, LENX, N, XOUT)
      EXTERNAL F
      DIMENSION Y(50)
        FUNCTION ALINE(DIST)
        DO 2012 I=1,LENX
 2012   Y(I) = X(I)+DIST*XDIR(I)
        ALINE = F(Y)
        END
      BESTD = ARGMAX(ALINE, 0.0, N)
      DO 2013 I=1,LENX
 2013 XOUT(I) = X(I)+BESTD*XDIR(I)
      END
\end{lstlisting}
Here we are using a library univariate maximizer to maximize the
univariate function \lstinline{ALINE}, which maps the distance along
the given direction to the value of our multidimensional function of
interest \lstinline{F} at that point.
Note that \lstinline{ALINE} refers to variables defined in its
enclosing scope, namely \lstinline{F}, \lstinline{X}, \lstinline{XDIR},
\lstinline{LENX}, and \lstinline{Y}.
Note that if \lstinline{ARGMAX} uses derivative information,
AD will be performed automatically on \lstinline{ALINE}.

\section{Concrete Example}\label{PaperXXsec:example}

We employ a concrete example to show the convenience of
the above constructs.
We will also illustrate the implementation on this example in
\cite{PaperXXimplementation}.
Let two companies, Apple and Banana, be engaged in
competition in a common fashion accessories market.  Each chooses a
quantity of their respective good to produce, and sells all produced
units at a price determined by consumer demand.  Let us model the goods as
being distinct, but partial substitutes, so that availability of
products of A decreases demand for products of B and vice versa (though
perhaps not the same amount).  We model both companies as having
market power, so the price each gets will depend on both their own
output and their competitor's.  Each company faces (different)
production costs and seeks to maximize its profit, so we can model
this situation as a two player game.  Let us further assume that
the quantities involved are large enough that discretization
effects can be disregarded.

An equilibrium $(a^*,b^*)$ of a two-player
game with continuous scalar strategies $a$ and $b$ and payoff
functions $A(a,b)$ and $B(a,b)$ must satisfy a system of equations:
\begin{align} \label{PaperXXeqn:ab}
  a^* &= \argmax_a A(a,b^*) &
  b^* &= \argmax_b B(a^*,b)
\end{align}
Equilibria can be sought by finding roots of
\begin{equation}\label{PaperXXeqn:game}
  a^* = \argmax_a A(a,\argmax_b B(a^*,b))
\end{equation}
which is the technique we shall employ.\footnotemark{}
Translated into computer code in the most natural way, solving this
equation involves
a call to an optimization subprogram within the function passed to an
optimization subprogram, itself within the function passed to a
root-finding subprogram.
If said optimization and root-finding subprograms need derivative
information, this gives rise to deeply nested AD.

\footnotetext{
The existence or uniqueness of an equilibrium is not in general
guaranteed, but our particular $A$ and $B$ have a unique equilibrium.
Coordinate descent (alternating optimization of $a^*$ and $b^*$)
would require less nesting, but has inferior convergence properties.
Although this example involves AD through iterative processes, we do
not address that issue in this work: it is beyond the scope of this
paper, and used here only in a benign fashion, for vividness.
}

Note that in \refeqn{PaperXXeqn:game}, the payoff function $B$ is
bivariate but $\argmax$ takes a univariate (in the variable of
maximization) objective function.
The $a^*$ variable passed to $B$ is \emph{free} in the innermost
$\argmax$ expression.  Free variables occur naturally in mathematical
notation, and we support them by allowing nested subprogram
definitions.

We can use our extensions to code finding the roots of
\refeqn{PaperXXeqn:game} in a natural style:
\begin{lstlisting}
C     ASTAR & BSTAR: GUESSES IN, OPTIMIZED VALUES OUT
      SUBROUTINE EQLBRM(BIGA, BIGB, ASTAR, BSTAR, N)
      EXTERNAL BIGA, BIGB
        FUNCTION F(ASTAR)
          FUNCTION G(A)
            FUNCTION H(B)
            H = BIGB(ASTAR, B)
            END
          BSTAR = ARGMAX(H, BSTAR, N)
          G = BIGA(A, BSTAR)
          END
        F = ARGMAX(G, ASTAR, N)-ASTAR
        END
      ASTAR = ROOT(F, ASTAR, N)
      END
\end{lstlisting}
where we implement just the minimal cores of one-dimensional
optimization and root finding to illustrate the essential point ---
root finding by the Rhapson method:
\begin{lstlisting}
      FUNCTION ROOT(F, X0, N)
      X = X0
      DO 1669 I=1,N
      CALL DERIV2(F, X, Y, YPRIME)
 1669 X = X-Y/YPRIME
      ROOT = X
      END
\end{lstlisting}
\begin{lstlisting}
      SUBROUTINE DERIV2(F, X, Y, YPRIME)
      EXTERNAL F
      ADF(X)
      Y = F(X)
      END ADF(YPRIME = TANGENT(Y))
      END
\end{lstlisting}
and optimization by finding the root of the derivative:
\begin{lstlisting}
      FUNCTION ARGMAX(F, X0, N)
        FUNCTION FPRIME(X)
        FPRIME = DERIV1(F, X)
        END
      ARGMAX = ROOT(FPRIME, X0, N)
      END
\end{lstlisting}
On our concrete objective functions these converge rapidly, so for
clarity we skip the clutter of convergence detection.

This strategy impels us to compute derivatives
nested five deep, in a more complicated pattern than just
a fifth-order derivative of a single function.
This undertaking is nontrivial with current AD
tools \cite{Siskind2008PtA}, but becomes straightforward with
the proposed extensions---embedded AD syntax and nested subprograms
make it straightforward to code sophisticated methods that require
complex patterns of derivative information.

\section{Discussion}\label{PaperXXsec:discuss}

The \PaperXXlang\ AD extensions hew to the spirit of \PaperXXfort, which
tends to prefer blocks rather than higher-order operators
for semantic constructs.
(In this, these extensions are syntactically quite similar to a set of
AD extensions integrated with the NAGware \PaperXXfort\ 95 compiler
\cite{Naumann2005CAw}, albeit quite different semantically.
Unfortunately those NAGware extensions are no longer publicly
available, and the limited available documentation is insufficient to
allow a detailed comparison.)
A reasonably straightforward implementation technology involves changing
transformed blocks into subprograms that capture their lexical
variable context and closure-converting these into top-level
subprograms, rendering them amenable to processing with existing tools
\cite{PaperXXimplementation}.
Since the machinery for nested subprograms is present, allowing
them imposes little additional
implementation effort.
Moreover, as seen in the example above, that extension makes
code that involves heavy use of higher-order functions, which is
encouraged by the availability of the AD constructs, more straightforward.
In this sense AD blocks and nested subprograms
interact synergistically.

These new constructs are quite expressive, but this very
expressiveness can tax many implementations, which might not support
some combinations or usages.
For instance, code which makes resolution of the AD at compile time
impossible (an $n$-th derivative subprogram, say)
would be impossible to support without a dynamic run-time AD
mechanism.
This would typically not be available.
Another common restriction would be that many tools do not support
reverse mode at all and even those that do typically do not allow
nesting over reverse mode, either reverse-over-reverse or forward-over-reverse.
It is the responsibility of the implementation to reject such
programs with a cogent error.

The \PaperXXlang\ extensions are implemented by
the \PaperXXpreprocessor\ preprocessor
\cite{PaperXXimplementation}, which generates
input for and invokes existing AD tools.
This leverages existing AD systems to provide the differentiation
functionality in a uniform and integrated way, extending the reach of
AD by making its use easier, more natural, and more widely applicable.

Such a prepreprocessor can target different AD systems (like
\PaperXXadifor\ \cite{Bischof1992AGD} and \PaperXXtapenade\
\cite{Hascoet2004TUG}), allowing easy porting of code from one AD system to
another.  It could even mix AD systems, for example using \PaperXXtapenade\ to
reverse-transform code generated by using \PaperXXadifor\ in forward mode,
capturing their respective advantages for the application at hand.  The effort
of implementing such retargetings and mixings could then be factored to one
developer (of the prepreprocessor) instead of many end users of AD.

A more important benefit of extending \PaperXXfort\ with AD syntax and nested
subprograms is that a host of notions become reusable
abstractions---not just first-order derivatives, but also their
variations, combinations, and uses, e.g.,
\begin{inparaitem}[]
\item Jacobians,
\item Hessians,
\item Hessian-vector products,
\item filters,
\item edge detectors,
\item Fourier transforms,
\item convolutions,
\item Hamiltonians,
\item optimizations,
\item integration,
\item differential-equation solvers.
\end{inparaitem}
The interfaces to different methods for these tasks can be made much
more uniform because, as our \lstinline{ARGMAX} did, they can accept subprograms
that just accept the variables of interest (in this case, the argument
of maximization) and take any needed side information from their
lexical scope; and subprograms such as \lstinline{ARGMAX} can obtain any
derivative information they wish from AD without having to demand it be passed
in as arguments.
So different maximization methods can be tried out on
the same objective function with ease, regardless of how much
derivative information they require; and at the same time, different
objective functions, that carry different side information, can be
maximized by the same maximization subprogram without having to adjust it
to transmit the needed side information.
Essentially, derivatives are requested where they are needed, and the
implementation does the necessary bookkeeping.

These modularity benefits are illustrated by our example program: the
\PaperXXlang\ input is only 64 lines of code, whereas the amount of code it
expands into, which is comparable to what would need to be written by hand
without these extensions, weighs in at a much more substantial 160 for
\PaperXXtapenade\ and 315 for \PaperXXadifor, including the configuration data
needed to run the AD preprocessors to produce the needed derivatives.
Manually performing the 5 nested applications of AD this example calls for
is a tedious, error prone, multi-hour effort, which must be undertaken
separately for each preprocessor one wishes to try.\footnote{A detailed
  step-by-step discussion of the transformation of
  this example along with all intermediate code is available at
  \protect\url{http://www.bcl.hamilton.ie/~qobi/fortran/}.}
Existing AD tools do already save the major labor
of manually implementing derivative and gradient subprograms, and keeping them
in sync with the subprograms being differentiated.  The further preprocessing
step outlined above leverages these tools into being even more useful.
For larger programs, the savings of implementation and maintenance effort would
be considerable.

The present suggestion is not, of course, limited to \PaperXXfort{}.
Nested subprograms have gained wide adoption
in programming-language designs from \PaperXXalgol~60 and beyond, and have
yielded proven gains in programmer productivity.  Their advantages for
code expressiveness have led to functions with lexical
scope being used as a mathematical formalism for reasoning about
computing \cite{PaperXXChurch41}, to programming languages organized around the
function as the primary program construct \cite{jones2003report}, and to
compilers that specialize in the efficient representation and use of
functions \cite{jones1993glasgow,steele1978rabbit}.


One can also
add \lstinline{ADF}- and \lstinline{ADR}-like constructs to other languages
that have preprocessor implementations of AD, for example, \PaperXXClang{} and
\PaperXXadic\ \cite{Bischof1997AAE}.  One would not even need to add nested
subprograms in the preprocessor, because \PaperXXGCC{} already implements them
for \PaperXXGNU~\PaperXXClang{}.
Doing so would expand the convenience (and therefore reach) of
existing AD technology even further.

\begin{figure}[tb]
\sidecaption
\psfrag{CPU Time (seconds)}[bc][bc]{\textbf{CPU Time (seconds)}}
\psfrag{Tapenade}[tc][tc]{\textbf{\PaperXXtapenade}}
\psfrag{Adifor}[tc][tc]{\textbf{\PaperXXadifor}}
\psfrag{Stalingrad}[tc][tc]{\textbf{\PaperXXStalingrad}}
\psfrag{0}[r][r]{0 }
\psfrag{5}[r][r]{5 }
\psfrag{10}[r][r]{10 }
\psfrag{6.97}[bc][bc]{6.97}
\psfrag{8.92}[bc][bc]{8.92}
\psfrag{5.83}[bc][bc]{5.83}
\includegraphics[width=0.55\textwidth]{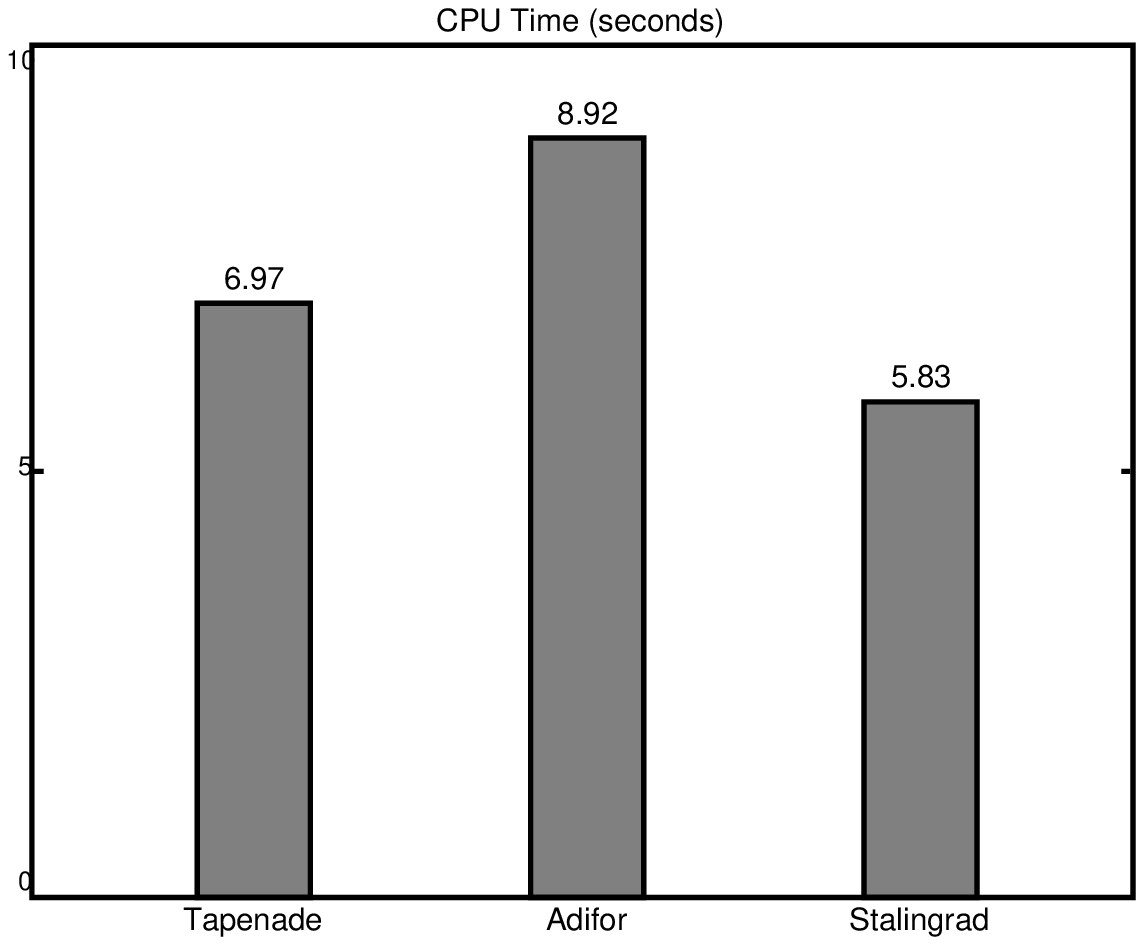}
\caption{\textbf{Performance comparison.}  Smaller is faster.
  Numeric solution of (\ref{PaperXXeqn:game}) with above
  \PaperXXlang\ code, $N=1000$ iterations at each level,
  \PaperXXpreprocessor\ targeting two \PaperXXfort-based AD
  tools; for comparison, the same computation is coded in \PaperXXVLAD\
  \cite{Pearlmutter2008UPL} and
  compiled with \PaperXXStalingrad\ \cite{Siskind2008UPU}.
  Computer: Intel i7 870 @ 2.93GHz, \textsc{Gfortran} 4.6.2-9, 64-bit Debian
  sid, \texttt{-Ofast} \texttt{-fwhole-program}, single precision.
  See \cite{PaperXXimplementation} for details.}
\label{PaperXXfig:benchmarks}
\end{figure}

Retrofitting AD onto existing languages by preprocessing is not
without its limitations, however.  Efficient AD preprocessors must
construct a call graph in order to determine which subprograms to
transform, along with a variety of other tasks already performed by
the compiler.
Moreover, optimizing compilers cannot be relied upon to aggressively
optimize intricate machine-generated code, as such code often exceeds
heuristic cutoffs in various optimization transformations.
This imposes a surprisingly serious limitation on AD preprocessors.
(Together, these also imply a significant duplication of effort, while
providing room for semantic disagreements between AD preprocessors and
compilers which can lead to subtle bugs.)
This leads us to anticipate considerable
performance gains from designing an optimizing compiler with
integrated AD.\@
Indeed, translating our concrete example into
\PaperXXVLAD\ \cite{Pearlmutter2008UPL} and compiling with
\PaperXXStalingrad\ \cite{Siskind2008UPU}, our prototype AD-enabled
compiler, justifies that suspicion (see \reffig{PaperXXfig:benchmarks}).
We therefore plan to make a
\PaperXXVLAD\ back-end available in version 2 of \PaperXXpreprocessor.

\section{Conclusion}\label{PaperXXsec:conclusion}

We have defined and motivated extensions to \PaperXXfort\ for convenient,
modular programming using automatic differentiation.
The extensions can be implemented as a prepreprocessor
\cite{PaperXXimplementation}.
This strategy enables modular, flexible use of AD in the
context of an existing legacy language and tool chain, without
sacrificing the desirable performance characteristics of these tools:
only about 20\%--50\% slower than a
dedicated AD-enabled compiler, depending on which \PaperXXfort\ AD
system is used.

\begin{acknowledgement}
This work was supported, in part, by Science Foundation Ireland grant
09/IN.1/I2637, National Science Foundation grant CCF-0438806, the Naval Research Laboratory under
Contract Number N00173-10-1-G023, and the Army Research Laboratory
accomplished under Cooperative Agreement Number W911NF-10-2-0060.
Any views, opinions, findings, conclusions, or recommendations contained or
expressed in this document or material are those of the author(s) and do not
necessarily reflect or represent the views or official policies, either
expressed or implied, of SFI, NSF, NRL, the Office of Naval Research, ARL, or
the Irish or U.S. Governments.
The U.S. Government is authorized to reproduce and distribute reprints for
Government purposes, notwithstanding any copyright notation herein.
\end{acknowledgement}

\bibliographystyle{plainnat}
\bibliography{../ad,../Papers/PaperXX/paper}
